\documentclass[conference]{IEEEtran}
\usepackage{eso-pic}
\usepackage{xcolor}
\definecolor{acceptblue}{RGB}{0,51,102}
\usepackage{cite}
\usepackage{amsmath,amssymb}
\usepackage{booktabs}
\usepackage{tabularx}
\usepackage{array}
\usepackage{xcolor}
\usepackage{url}
\usepackage{graphicx}
\usepackage{microtype}
\usepackage{tikz}
\usetikzlibrary{arrows.meta,positioning,fit,shapes.multipart}
\usepackage{enumitem}
\usepackage{balance}
\usepackage[final,hidelinks,bookmarks=false]{hyperref}
\hypersetup{
  pdftitle={A2A-ForensicTrace: Offline Verification of Tamper-Evident A2A Runtime Evidence},
  pdfauthor={Adil Alshammari, Sareh Assiri, Hayretdin Bahsi},
  pdfkeywords={Agent2Agent protocol, A2A security, digital forensics, tamper-evident evidence, incident trace, offline verification}
}

\newcommand{\code}[1]{\texttt{\detokenize{#1}}}
\newcommand{\decisioncode}[2]{\begin{tabular}[t]{@{}l@{}}\code{#1_}\\\code{#2}\end{tabular}}

\newcolumntype{Y}{>{\raggedright\arraybackslash}X}
\newcolumntype{L}[1]{>{\raggedright\arraybackslash}p{#1}}

\title{A2A-ForensicTrace: Offline Verification of Tamper-Evident A2A Runtime Evidence}

\author{
\IEEEauthorblockN{
Adil Alshammari\textsuperscript{1,2}
\quad
Sareh Assiri\textsuperscript{3}
\quad
Hayretdin Bahsi\textsuperscript{1,4}
}%
\IEEEauthorblockA{
\textsuperscript{1}School of Informatics, Computing, and Cyber Systems,
Northern Arizona University, Flagstaff, Arizona, USA\\
\textsuperscript{2}College of Computer and Information Sciences,
Majmaah University, Majmaah 11952, Saudi Arabia\\
\textsuperscript{3}College of Engineering and Computer Science,
Jazan University, Jazan, Saudi Arabia\\
\textsuperscript{4}School of Information Technologies,
Tallinn University of Technology, Tallinn, Estonia\\
aha388@nau.edu;
sassiri@jazanu.edu.sa;
Hayretdin.Bahsi@nau.edu
}
}

\begin{document}
\AddToShipoutPictureFG*{%
  \AtPageUpperLeft{%
    \raisebox{-12mm}{%
      \makebox[\paperwidth][c]{%
        \normalfont\footnotesize
        \textcolor{acceptblue}{%
          This paper was accepted to IEEE IEMCON 2026%
        }%
      }%
    }%
  }%
}

\maketitle
\begin{abstract}
Security-relevant Agent2Agent (A2A) executions can cross organizational boundaries, leaving investigators without live access to all participating systems. Offline investigation involves checking preserved records and their cross-record relationships for consistency. This paper presents
\emph{A2A-ForensicTrace}, an offline verification layer that converts runtime observations into typed records, derives protocol-relevant relationships, and commits both under an incident-trace root. An Ed25519-signed receipt binds the
root and capture digest to the incident context. Evaluation comprised 240 executions through the official A2A software development kit (SDK), with actions selected by a large language model (LLM). These included 120 condition runs and 120 matched controls. All condition runs returned the expected bounded findings. No control produced an indication, and all roots and receipts verified. Median in-memory latency of the full offline verifier was 1.61~ms for the scenario traces. In the separate scaling experiment, it was 30.85~ms at 1,000 committed leaves. Future work will broaden A2A lifecycle coverage.
\end{abstract}

\begin{IEEEkeywords}
Agent2Agent protocol, A2A security, digital forensics, tamper-evident traces, incident trace, offline verification.
\end{IEEEkeywords}

\section{Introduction}
Version 1.0 of the Agent2Agent (A2A) protocol specification defines an interoperability layer through which independent agents can discover capabilities, coordinate tasks, exchange artifacts, and stream updates. It defines the \emph{AgentCard}, task and context identifiers, task states, artifacts, streaming, push notifications, and protocol bindings~\cite{noauthor_a2a_nodate}. When preserved, selected instances of these runtime objects can support post-incident digital forensic investigation by helping reconstruct a security-relevant execution~\cite{lyle_digital_2022}.

Consider an agent that discovers a remote peer through an AgentCard, sends a JSON-RPC \code{message/send} request to the selected endpoint, receives a task, and later obtains an artifact. If a preserved identifier, lifecycle transition, endpoint binding, or artifact digest is inconsistent, the final output alone cannot show where the inconsistency arose. Therefore, a forensic investigation needs the selected records and the relationships that connect them. For later investigation, this evidence should be retained in append-only or otherwise immutable storage and protected so that insertion, deletion, modification, or reordering is detectable~\cite{crosby_efficient_2009,nelson_incident_2025}. A2A-ForensicTrace uses cryptographic commitments, a whole-capture digest, and a signed receipt to make tampering detectable. The prototype does not implement the evidence repository or its access and retention controls.

Prior work studies A2A security, deployment controls, sensitive-data handling, message authentication, and execution provenance~\cite{narajala_building_2025,louck_improving_2025,anbiaee_security_2026,alshammari_authenticated_2026,wang_agent_2026}. To our knowledge, these works do not define an integrity-gated, multi-record A2A incident trace that commits protocol relationships and supports reproducible offline forensic interpretation. This gap motivates the following research question.

\noindent\textbf{RQ:} How can selected A2A runtime observations be preserved as a tamper-evident incident trace, verified for integrity, and interpreted offline for bounded protocol-level inconsistencies?

This paper presents \emph{A2A-ForensicTrace}, a post-incident evidence and verification layer. Its forensic unit is a \emph{multi-record incident trace}, which contains preserved records and explicitly committed relationships for interpretation under the selected verifier profile. 

A \emph{forensic verification profile} is a versioned configuration defined by A2A-ForensicTrace. It specifies the records and relationships required for an investigative question, the enabled checks, the admissibility conditions for interpretation, and the decision rules. For example, selected-card analysis requires discovery, AgentCard, and endpoint records. Analysis of uniform resource identifier (URI) retrieval requires URI reference, dereference, and response records. In this evaluation, a \emph{condition family} adapts one of the six attack families in the recent A2ASecBench study and contains the condition runs and matched benign controls used to evaluate one bounded forensic rule~\cite{li_a2asecbench_2026}. We evaluate one profile version using six family-specific manifests. The three verifier settings characterize integrity verification and forensic interpretation within the implemented profile, while a separate experiment measures verifier scaling.

The implementation uses the official A2A software development kit (SDK) execution path and preserves selected fields from SDK-native objects together with profile-specific testbed observations for offline verification~\cite{google_llc_a2a-sdk_2026}.

To answer this RQ, the contributions are:
\begin{itemize}[leftmargin=*]
    \item an A2A runtime evidence model that maps selected SDK fields and testbed observations to typed forensic records;
    \item domain-separated commitments for records and deterministically derived cross-record relationships, summarized by an incident trace root and bound to incident context by an Ed25519-signed receipt;
    \item an offline verifier that separates trace-integrity validation from protocol-aware interpretation and returns structured, reason-coded findings; and
  \item an official-SDK evaluation of 240 condition and matched-control executions with actions selected by a large language model (LLM), supplemented by a three-configuration comparison and a 3,500-run verifier-scaling experiment.
\end{itemize}

This work addresses a different verification layer from application-level case-evidence verification. It does not determine whether business events support a case outcome. Instead, it verifies the integrity and profile-bounded consistency of protocol runtime evidence, including AgentCard selection, exchanges, task transitions, artifacts, and their typed relationships.

\section{Related Work}
\label{sec:relatedwork}
The A2A specification and official SDK define the objects and execution path from which runtime evidence can be captured~\cite{noauthor_a2a_nodate,google_llc_a2a-sdk_2026}. JSON-RPC provides request--response correlation through its \code{id} field~\cite{json-rpc_working_group_json-rpc_2013}. The A2A specification and SDK do not define a tamper-evident incident-trace model or an offline forensic verifier.

Authenticated A2A messaging verifies an individual preserved message and its integrity evidence~\cite{alshammari_authenticated_2026}. A2A-ForensicTrace instead verifies a multi-record trace and the cross-record relationships needed for protocol-aware interpretation.

A2ASecBench is the closest protocol-specific security framework~\cite{li_a2asecbench_2026}. It evaluates six attack families with matched benign controls and reports attack success and utility. We use its taxonomy only to select evidence surfaces and matched controls. Our verifier asks whether preserved evidence supports a bounded finding. The harness outcome is withheld until verification is complete. Other studies address deployment controls, sensitive data, or threat modeling~\cite{narajala_building_2025,louck_improving_2025,anbiaee_security_2026}.

Evidence-tracing research models agent runs through steps, outputs, actions, and relations~\cite{wang_agent_2026}, while W3C PROV provides a domain-independent provenance model~\cite{moreau_prov-dm_2013}. Software-supply-chain systems such as in-toto and SCITT bind statements or attestations to artifacts and transparency services~\cite{torres-arias_-toto_2019,h_birkholz_architecture_2026}. These approaches do not define the evaluated A2A record groups, derived relationship requirements, or reason-coded decision semantics.

Tamper-evident logging makes unauthorized changes to an append-only history detectable through authenticated commitments and consistency auditing~\cite{crosby_efficient_2009}. Our unit instead commits typed A2A records and their derived relationships for profile-bound interpretation. The integrity layer builds on independently checkable roots and deterministic representations~\cite{laurie_certificate_2021,rundgren_json_2020}, while incident-response guidance motivates evidence preservation~\cite{nelson_incident_2025}. The contribution is therefore the A2A-specific evidence/relationship profile, integrity-gated semantics, and evaluation---not a new hash tree or signature scheme. Table~\ref{tab:positioning} summarizes the distinction.

\begin{table}[t]
\centering
\caption{Comparison with the closest problem formulations.}
\label{tab:positioning}
\scriptsize
\setlength{\tabcolsep}{2pt}
\begin{tabularx}{\columnwidth}{L{0.215\columnwidth} L{0.215\columnwidth} L{0.25\columnwidth} Y}
\toprule
Work & Unit of analysis & Primary objective & Primary output \\
\midrule
A2ASecBench~\cite{li_a2asecbench_2026} & System execution & Security benchmark: attack/utility evaluation & Attack success rate and utility \\
A2A message authentication~\cite{alshammari_authenticated_2026} & Individual message & Message integrity & Validity result \\
Agent-provenance survey~\cite{wang_agent_2026} & Execution trace/graph & Lineage methods & Taxonomy/design space \\
A2A-ForensicTrace & A2A incident trace & Integrity plus bounded interpretation & Reason-coded report \\
\bottomrule
\end{tabularx}
\end{table}

\section{A2A Runtime Incident Model}
\label{sec:incidentmodel}
The evaluated verifier profile covers four runtime evidence surfaces and post-capture integrity. The verifier receives a trace bundle, a signed incident receipt, the selected verifier profile, and an external signer--key binding.

An \emph{observation} is a value or event seen during execution. A \emph{record} is one typed item preserved from an observation. It may contain selected fields from an A2A object or a testbed measurement and is not necessarily an original log entry. A \emph{relationship}, also called a forensic edge, is a link derived between two records from their preserved fields.

An \emph{incident trace} consists of trace metadata, selected records, and their derived relationships. In the evaluated workflow, it begins with recorded AgentCard discovery and closes after the final SDK response and all profile-required observations have been captured. Its metadata records the trace and run identifiers, profile version, capture source, and capture time.

\subsection{Evidence Scope and Trust Assumptions}
\textbf{Assets and verifier input.} The protected object is the finalized trace bundle containing typed records, trace metadata, and relationships derived from preserved fields. A versioned Ed25519 receipt binds its incident root and whole-capture digest to the incident, profile, capture context, and signer. The verifier resolves the signer and key against an external trust store.

\textbf{Adversary and trusted base.} After signing, an adversary may read the bundle and receipt and may insert, remove, or alter records, relationships, runtime-order fields, or signed receipt fields. The adversary is assumed unable to find SHA-256 collisions, forge Ed25519 signatures under a trusted key, or exploit ambiguity in the restricted deterministic serializer. The capture adapter, receipt signer, verifier, and external trust store form the trusted computing base.

\textbf{Claimed property.} Under these assumptions, post-signing changes to the finalized bundle or signed context fail an integrity check, except with negligible cryptographic probability. Record and edge list order is normalized when constructing the incident root, whereas the signed whole-capture digest also binds the finalized list representation. Logical runtime order is interpreted only from committed timestamps, sequence numbers, and explicit predecessor or correlation fields. Pre-capture omissions and trusted-component compromise remain outside this property (Section~\ref{sec:discussion}).

\subsection{Runtime Evidence Surfaces}
\textbf{Discovery and capability use.} The A2A specification defines the AgentCard. The terms \emph{candidate}, \emph{trusted}, and \emph{selected} describe roles in our evidence schema, not separate A2A object types. A candidate card was considered during discovery. A trusted card is a reference from the configured trust set. The selected card supplied the endpoint used for the request.

\textbf{SDK exchanges and task evidence.} An \emph{exchange} is one JSON-RPC request and response pair correlated by the \code{id} field. A capability invocation records the capability observed during that exchange.

\textbf{Routing and task load.} A routing event records one observed agent-to-agent step, including its source, destination, and whether progress occurred. A task-state transition records one change in task state. A task-load snapshot records the number of tasks awaiting further input together with queue-depth, latency, or timeout measurements. The saturation rule requires both a configured task-count threshold and evidence of service degradation.

\textbf{Post-capture trace integrity.} The receipt signature, signed context, capture digest, typed record commitments, and relationship commitments are verified before protocol-aware interpretation.

\textbf{URI and artifact evidence.} 
URI records show where a URI appeared, whether retrieval was attempted, the resolved destination, and any response. The selected profile supplies the allowed-destination policy. These records support the evaluated agent-side request forgery (ASRF) rule. Artifact records preserve received content and media type. A render record states that a renderer received the artifact, while a sanitizer record states whether sanitization was applied. In the evaluated artifact-triggered script injection (ATSI) condition, an \emph{active-content indicator} is an inert script-like pattern used only to exercise the rule.

\textbf{Record sources.} AgentCard snapshots, JSON-RPC requests and responses, and returned task, context, state, and artifact fields come from the SDK-facing execution. The capture harness records card roles, observed capability use, routing steps, task-load measurements, URI retrieval and response, artifact rendering, and sanitization around that exchange. The selected profile supplies trusted-card references, thresholds, and URI policy. Here, \emph{official-SDK} identifies the exercised protocol path. The A2A-ForensicTrace schema combines SDK-facing fields with linked, profile-specific observations recorded by the capture harness.

\begin{figure*}[!t]
\centering
\begin{tikzpicture}[
  >=Latex,
  node distance=5mm,
  sdk/.style={draw, rounded corners=1.5pt, align=center, fill=blue!7,
    minimum height=13mm, text width=36.5mm,
    font=\fontsize{9}{11}\selectfont, inner sep=3pt},
  evidence/.style={sdk, fill=green!7},
  commit/.style={sdk, fill=violet!8},
  report/.style={sdk, fill=purple!7},
  group/.style={draw=black!60, dashed, rounded corners=2pt, inner sep=5pt},
  arrow/.style={-{Stealth[length=2mm]}, thick}
]
\node[sdk] (card) {1. \textbf{AgentCard}\\well-known endpoint};
\node[sdk, right=of card] (rpc) {2. \textbf{JSON-RPC}\\\code{message/send}};
\node[sdk, right=of rpc] (exec) {3. \textbf{SDK execution}\\\code{AgentExecutor}\\\code{RequestContext}};
\node[sdk, right=of exec] (task) {4. \textbf{Task evidence}\\state + artifact};

\node[evidence, below=11mm of card] (capture) {5. \textbf{Runtime capture}\\adapter};
\node[evidence, right=of capture] (typed) {6. \textbf{Typed evidence}\\records + derived edges};
\node[commit, right=of typed] (root) {7. \textbf{Incident trace root}\\+ Ed25519 receipt};
\node[report, right=of root] (verify) {8. \textbf{Offline verifier}\\reason-coded report};

\draw[arrow] (card) -- (rpc);
\draw[arrow] (rpc) -- (exec);
\draw[arrow] (exec) -- (task);
\draw[arrow] (task.south) -- ++(0,-5mm) -| (capture.north);
\draw[arrow] (capture) -- (typed);
\draw[arrow] (typed) -- (root);
\draw[arrow] (root) -- (verify);

\node[group, fit=(card)(rpc)(exec)(task),
  label={[font=\fontsize{9}{11}\selectfont\bfseries]above:Official A2A SDK execution path}] {};
\node[group, fit=(capture)(typed)(root)(verify),
  label={[font=\fontsize{9}{11}\selectfont\bfseries]below:
  A2A-ForensicTrace evidence capture and offline verification}] {};
\end{tikzpicture}
\caption{A2A-ForensicTrace architecture from the official A2A SDK execution
path to committed typed evidence and reason-coded offline verification.}
\label{fig:model}
\end{figure*}
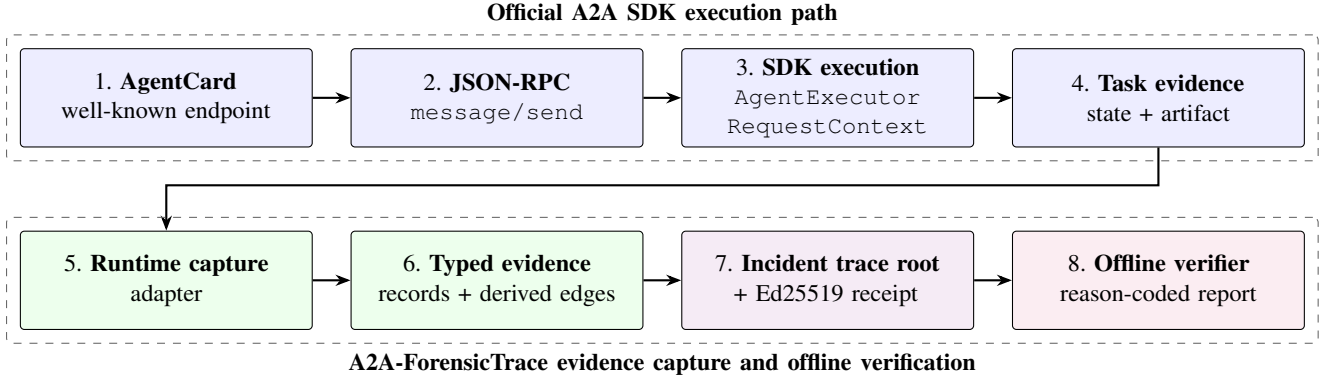

Fig.~\ref{fig:model} summarizes the two-stage path. Steps 1--4 show the official A2A SDK execution. Steps 5--7 capture, structure, and commit the resulting evidence. Step 8 verifies it offline.

\section{A2A-ForensicTrace Design}
\label{sec:design}
A2A-ForensicTrace operates in four stages. First, the runtime capture adapter selects SDK fields and linked testbed observations. Second, the incident trace schema represents each observation as a typed record and derives relationships between records. Third, the commitment layer hashes every record and relationship, constructs the incident trace root, and creates an authenticated receipt. Fourth, the offline verifier checks integrity and then applies the selected profile. It produces a machine-readable report containing a decision, reason codes, and affected-record identifiers.

\subsection{Runtime Forensic Records}
The following names identify record types defined by the A2A-ForensicTrace schema. Each record captures selected fields from an SDK-facing A2A object or a linked testbed observation. \textbf{TraceMetadata} describes the trace boundary and capture context. Discovery evidence uses \textbf{AgentCardSnapshot}, \textbf{SelectedEndpoint}, and \textbf{DiscoveryEventEvidence}. Exchange evidence uses \textbf{RuntimeExchange} and \textbf{CapabilityInvocationEvidence}. Routing and task evidence use \textbf{RoutingEventEvidence}, \textbf{TaskStateTransition}, and \textbf{TaskLoadSnapshot}. URI evidence uses \textbf{UriReferenceEvidence}, \textbf{UriDereferenceEvidence}, and \textbf{DereferenceResponseEvidence}. Artifact handling uses \textbf{ArtifactEvidence}, \textbf{ArtifactRenderEvidence}, and \textbf{SanitizerEvidence}.

\subsection{Derived Relationships}
A forensic relationship links two preserved records deterministically. For example, \code{DISCOVERY_}\allowbreak\code{SELECTED_}\allowbreak\code{CARD} links a discovery event to the selected AgentCard. The relation \code{AGENTCARD_}\allowbreak\code{RESOLVES_}\allowbreak\code{TO_}\allowbreak\code{ENDPOINT} links that card to its endpoint. Similarly, \code{EXCHANGE_}\allowbreak\code{INVOKED_}\allowbreak\code{CAPABILITY} links an exchange to its observed capability invocation. Other relations connect routing steps, task-load slots, URI dereferences and responses, and artifact, render, and sanitizer records. These names belong to the A2A-ForensicTrace schema.

A derived relationship
$e=(\tau_s,i_s,\tau_d,i_d,\rho,m)$ links a source record to a destination record. Here, $\tau_s$ and $\tau_d$ are the source and destination record types, $i_s$ and $i_d$ are their identifiers, $\rho$ is the relation type, and $m$ is optional relation metadata. Thus, source record $(\tau_s,i_s)$ has relation $\rho$ to destination record $(\tau_d,i_d)$. The metadata $m$ contains only information needed to interpret the relation, such as a card-similarity score, source and destination agent identifiers, or an occupied-slot identifier. The edge envelope $E_e=(v,i_e,e)$ contains the profile version $v$, the derived edge identifier $i_e$, and the edge payload $e$. It is serialized and hashed to produce the relationship commitment.

\subsection{Typed Commitments and Incident Root}
The prototype serializes commitment inputs as UTF-8 JSON with sorted keys and compact separators. It rejects non-finite numbers. 
Let $H$ denote SHA-256, let $S(x)$ denote the UTF-8 byte string
produced by this restricted JSON serialization of $x$, and define
\[
D(t,z)=H(\mathrm{UTF8}(t)\,\|\,\mathtt{0x00}\,\|\,z),
\]
where $t$ is a domain-separation tag, $z$ is a serialized byte string, and $\|$ denotes byte-string concatenation. This serializer is deterministic for the evaluated data model but is not claimed to implement the complete JSON Canonicalization Scheme (JCS)~\cite{rundgren_json_2020}.

For one record $r_x$, its commitment input is the envelope
$E_x=(v,c_x,t_x,i_x,p_x)$. Here, $v$ is the profile version,
$c_x$ is the name of the record list in the bundle, $t_x$ is the record type, $i_x$ is its identifier, and $p_x$ is the complete preserved content of that record. For example, the collection name may be \code{exchanges}, while the payload contains that exchange's request and response fields. The capture adapter obtains $i_x$ from the required type-specific field, such as \code{exchange_id} or \code{artifact_id}. A missing identifier or conflicting reuse of an identifier is rejected.

When a rule produces a finding, the affected-record identifier in the report is the existing $i_x$ of the implicated record. The verifier does not create a second identifier.

For every evaluated record $r_x$ and derived relationship $e$, the record commitment $H_{r_x}$, relationship commitment $H_e$, and incident trace root $R_T$ are computed as

\begin{align}
H_{r_x} &= D(\mathtt{A2AFT.RecordEnvelope.v3},S(E_x)),\\
H_e &= D(\mathtt{A2AFT.EdgeEnvelope.v3},S(E_e)),\\
R_T &= \mathrm{MerkleRoot}(\Lambda_T),
\end{align}

where $\Lambda_T$ is the ordered sequence of all record and relationship commitments. The literal strings supplied as the first argument to $D$ are domain-separation tags. They ensure that record and relationship envelopes belong to distinct commitment domains, even if their serialized inputs are identical.

The fixed profile collection order starts with trace metadata, then candidate, trusted, and selected cards. Next come endpoints and discovery, followed by exchanges and capability invocations. Routing, task transitions, and task load follow. The final groups are URI reference, dereference, and response, then artifact, render, and sanitizer evidence. Records within each collection are sorted by their required identifier. Each derived relationship identifier is the first 24 hexadecimal characters of a domain-separated hash over its endpoints, relation type, and metadata. Derived relationships are sorted by this identifier. For $n$ records and $q$ edges,
\begin{equation}
\Lambda_T=(H_{r_1},\ldots,H_{r_n},H_{e_1},\ldots,H_{e_q}),
\end{equation}
and the already domain-separated $H_{r_x}$ and $H_e$ values are the Merkle leaves. Internal nodes are hashed with the \code{A2AFT.MerkleNode.v3} tag. When a level contains an odd number of nodes, its final node is duplicated so that every node has a pair before the next level is computed. The fixed collection order makes root reconstruction reproducible, while committed timestamps, sequence numbers, and relationship fields express runtime chronology.

Let $T$ denote the complete finalized trace bundle containing trace metadata, all record lists, and the preserved relationships. Its whole-capture digest is
\begin{equation}
C_T=D(\mathtt{A2AFT.Capture.v3},S(T)).
\end{equation}

The whole-capture digest $C_T$ is stored in the authenticated receipt together with $R_T$. The root $R_T$ commits to the deterministically ordered record and relationship envelopes. Its definition does not separately encode the leaf count $n+q$. The signed $C_T$ binds the complete serialized bundle, including list lengths and order, and is checked alongside $R_T$.

The Ed25519 signature covers $R_T$, $C_T$, the incident and profile identifiers, and hashes of the predefined family definitions and experiment settings. It also covers the capture backend, signer and key identifiers, signature algorithm, and issue time~\cite{josefsson_edwards-curve_2017}.

For each condition family, the signed hash binds its predefined condition and matched-control inputs, required evidence and relationships, thresholds, and expected external observation. The execution schedule is also defined before the experiment and records the run order. A separate signed hash binds settings shared by all runs, including the backend, model configuration, and schedule seed. The harness constructs a unique incident identifier from the run identifier. The profile identifier names the versioned schema and rule configuration. The configured signer identifier names the signing authority. The key identifier is derived from a SHA-256 fingerprint of the Ed25519 public key. Together, these values bind the receipt to one run, one profile, and one trusted verification key.

\section{Offline Incident Verifier}
\label{sec:verifier}
Verification begins with integrity. The verifier authenticates the receipt and checks its signed context. It recomputes and compares the whole-capture digest. It then recomputes the record and edge commitments, reconstructs the incident root, and compares it with the signed expected root. It then applies a second gate, called \emph{profile admissibility}. This gate checks schema validity, identifier and reference consistency, relationship completeness, and the profile-specific severity of missing evidence.

Only an intact trace reaches the \emph{profile-admissibility} gate. The verifier returns \code{TRACE_REJECTED} if a required record collection is malformed or the preserved-relationship list is absent. Missing or duplicate record or relationship identifiers, or relationships that refer to missing records, also cause rejection. The same applies when preserved relationships differ in type, endpoints, or multiplicity from those rederived from the committed records. This topology comparison does not compare edge identifiers or metadata with their rederived values. Both remain bound by the signed capture digest and incident root. Structurally valid but incomplete evidence returns \code{INCONCLUSIVE}.

The six enabled rules adapt the six attack families reported by the recent A2ASecBench study~\cite{li_a2asecbench_2026}. They provide a representative attack-informed basis for evaluating the offline verifier and are not claimed to constitute a complete set of A2A forensic analyses.
\begin{enumerate}[leftmargin=*]
    \item \textbf{AgentCard spoofing:} the selected card is highly similar to a trusted card but has different publisher provenance.
    \item \textbf{Capability cloaking:} an observed runtime capability is not declared by the selected AgentCard.
    \item \textbf{Cycle overflow:} routing contains a repeated directed cycle with no progress before the configured stopping bound.
    \item \textbf{Half-open task flooding:} tasks awaiting input reach the configured threshold while queue, latency, or timeout evidence also indicates service degradation.
    \item \textbf{Agent-side request forgery (ASRF):} an agent dereferences a destination disallowed by the selected profile and receives a response.
    \item \textbf{Artifact-triggered script injection (ATSI):} an artifact containing an inert active-content pattern reaches a renderer without recorded sanitization.
\end{enumerate}
Table~\ref{tab:taxonomy} maps these condition families to the required evidence and reason-coded findings.

In this paper, a preserved incident trace is the finalized bundle containing the captured runtime records, their derived relationships, and the associated integrity evidence. A \emph{root-only configuration} is an evaluation setting, not another forensic profile. It verifies only the receipt, whole-capture digest, and incident trace root. Its purpose is to confirm trace integrity without interpreting the record contents. A \emph{records-only configuration} checks the typed records but does not use the relationships listed in Table~\ref{tab:taxonomy} when applying the forensic rules. Because the evaluated profile requires relationship evidence, all 240 traces returned \code{INCONCLUSIVE} when relationship use was disabled. This comparison does not evaluate an alternative verifier that derives relationships from committed records without separate relationship commitments. The full configuration applies the same rules using both records and relationships.

The root-only configuration records \code{INTEGRITY_VERIFIED} as an internal status when all integrity checks pass.

\begin{table}[t]
\caption{Verifier decision semantics.}
\label{tab:decisions}
\centering
\small
\begin{tabular}{p{0.30\columnwidth}p{0.59\columnwidth}}
\toprule
Decision & Meaning \\
\midrule
\decisioncode{FORENSICALLY}{CONSISTENT} &
For the admitted evidence and selected profile, all required checks complete and no enabled indicator is supported. \\

\decisioncode{ANOMALY}{INDICATED} &
The admitted trace supports at least one enabled protocol-level indicator. \\

\code{INCONCLUSIVE} &
Integrity and admissibility pass, but the available evidence is insufficient
or ambiguous for the requested conclusion. \\

\decisioncode{TRACE}{REJECTED} &
The receipt signature or signed context is invalid, capture-digest or root verification fails, or a fatal schema, reference, or relationship check fails. Interpretation does not proceed. \\
\bottomrule
\end{tabular}
\end{table}

\code{ANOMALY_INDICATED} reports which preserved inconsistency satisfied an enabled rule, such as a policy-disallowed URI retrieval or unsanitized active artifact content. It is an offline evidence-based indication. It does not establish exploit success, intent, attribution, or completeness of the captured trace.

\section{Forensic Evaluation}
\label{sec:evaluation}
The evaluation separates a primary LLM-backed execution matrix from supporting offline-verifier tests. An \emph{evaluation matrix} is a fixed schedule of executions across predefined factors. A \emph{condition run} is a controlled A2A interaction containing the input or local state needed to exercise the corresponding attack-informed forensic rule. Its \emph{matched control} is the corresponding benign interaction, in which that property is removed or neutralized while the same task purpose, model, permitted action set, and SDK route are retained. Each family contains 20 condition runs and 20 matched controls.

Before execution, the condition and control inputs, local parameters, required evidence and relationships, thresholds, and expected external observation were fixed for each family. These values were not changed after the runs began. A fixed seed determined the run order. The receipt contains hashes of these family-specific values and the settings shared by all runs. The harness records the per-run external reference outcome separately and uses it only after verification to assess the report. The verifier receives the profile information needed to apply its rules, but not this per-run reference outcome. The model receives neither an explicit condition/control label nor the expected finding.

For each condition or control run, the pinned model selects one action permitted by the task schema. The harness checks that the selected action follows the schema and applies it before SDK execution. The offline verifier independently processes the resulting preserved bundle and receipt.

The policy wrapper makes one generation request per invocation and rejects invalid responses without automatic regeneration.

NIST IR 8354 distinguishes validation from verification~\cite{lyle_digital_2022}. Validation assesses whether a method is fit for its intended purpose. Verification checks whether an implementation follows its design. In our evaluation, we compare the verifier's decisions with the expected decisions under the evaluated conditions to assess whether the selected profile is fit for its stated purpose. The supporting tests check whether the implementation follows its design.

\subsection{Setup}

The evaluated implementation used \code{a2a-sdk==1.1.0} through its non-streaming JSON-RPC \code{message/send} compatibility path. The final evaluation contained 240 LLM-backed official-SDK executions. The experiments ran on a Microsoft Surface Laptop Studio~2 with Windows~11 Home (x64), an Intel Core i7-13700H processor, 32~GB RAM, and Python~3.13.3. Each execution used Ollama~0.32.4 and a digest-pinned \code{llama3.2:3b} model to select one schema-constrained action before the official-SDK call. One warm-up preceded the randomized execution schedule and was excluded from the reported data. All trial services ran on the same computer and were accessed through loopback network addresses, so no external destination was contacted. All executions used the same external Ed25519 signer--key binding.

Before the final schedule, a separate 24-run readiness pilot checked the end-to-end operation of model action selection, official-SDK execution, evidence capture, receipt generation, and offline verification. The pilot runs were excluded from all reported counts and statistics.

A scenario is one condition or matched-control execution within one of the six evaluated families. \emph{Scenario latency} is the in-memory execution time of the full offline verifier measured once for each of the 240 preserved LLM-backed traces after the bundle and signed receipt were loaded. \emph{Scaling latency} is measured separately on purpose-built valid bundles of increasing size to isolate how the number of record and relationship commitments affects verification time. Table~\ref{tab:params} summarizes both experiments.

\begin{table}[t]
\centering
\caption{Evaluation setup.}
\label{tab:params}
\footnotesize
\begin{tabularx}{\columnwidth}{L{0.38\columnwidth} Y}
\toprule
Parameter & Value \\
\midrule
LLM-backed official SDK & 240: 6 families $\times$ (20 condition + 20 matched control) \\
Three-configuration comparison & 240 LLM-backed traces $\times$ 3 configurations = 720 configuration-level results \\
Scenario latency & 240 full-verifier timings, one per LLM-backed trace \\
Verifier scaling & 7 commitment counts; 500 timings/count; 3,500 timed verifications \\
Commitment and receipt & Domain-separated SHA-256; Ed25519 trusted binding \\
\bottomrule
\end{tabularx}
\end{table}

\begin{figure*}[!t]
\centering
\begin{tikzpicture}[
  box/.style={
    draw, rounded corners=2pt, thick, align=center,
    minimum height=13mm, text width=35mm,
    font=\fontsize{9}{11}\selectfont, inner sep=3pt, outer sep=0pt
  },
  model/.style={box, fill=blue!8},
  sdk/.style={box, fill=green!8},
  evidence/.style={box, fill=orange!9},
  check/.style={box, fill=red!7},
  arrow/.style={-{Stealth[length=2.0mm]}, thick},
  group/.style={draw, dashed, rounded corners=2pt, inner sep=5pt}
]

\node[model] (prompt) at (0,0)
  {Frozen prompt\\and local setup};
\node[model] (model) at (4.25,0)
  {Pinned Ollama model\\schema-constrained action};

\node[sdk] (sdk) at (8.90,0)
  {Official A2A SDK\\\code{message/send}};
\node[sdk] (capture) at (13.15,0)
  {Runtime capture\\typed observations};

\node[evidence] (records) at (13.15,-2.20)
  {Records + relationships\\incident root + receipt};
\node[evidence] (verifier) at (8.90,-2.20)
  {Offline verifier\\integrity, then interpretation};
\node[check] (compare) at (4.25,-2.20)
  {Post-verification comparison\\decision + reason code + affected record};
\node[check] (outcome) at (0,-2.20)
  {External outcome check\\outside verifier input};

\draw[arrow] (prompt) -- (model);
\draw[arrow] (model) -- (sdk);
\draw[arrow] (sdk) -- (capture);
\draw[arrow] (capture) -- (records);
\draw[arrow] (records) -- (verifier);
\draw[arrow] (verifier) -- (compare);
\draw[arrow] (outcome) -- (compare);

\node[group, fit=(prompt)(model)] (modelgroup) {};
\node[font=\fontsize{9}{11}\selectfont, above=2pt of modelgroup]
  {Model action selection};

\node[group, fit=(sdk)(capture)(records)(verifier)] (sdkgroup) {};
\node[font=\fontsize{9}{11}\selectfont, below=2pt of sdkgroup]
  {SDK execution, evidence capture, and offline verification};

\end{tikzpicture}

\caption{LLM-backed A2A evaluation: official SDK execution of model-selected actions, with external reference outcomes used only to assess verifier reports.}

\label{fig:promptvalidation}
\end{figure*}
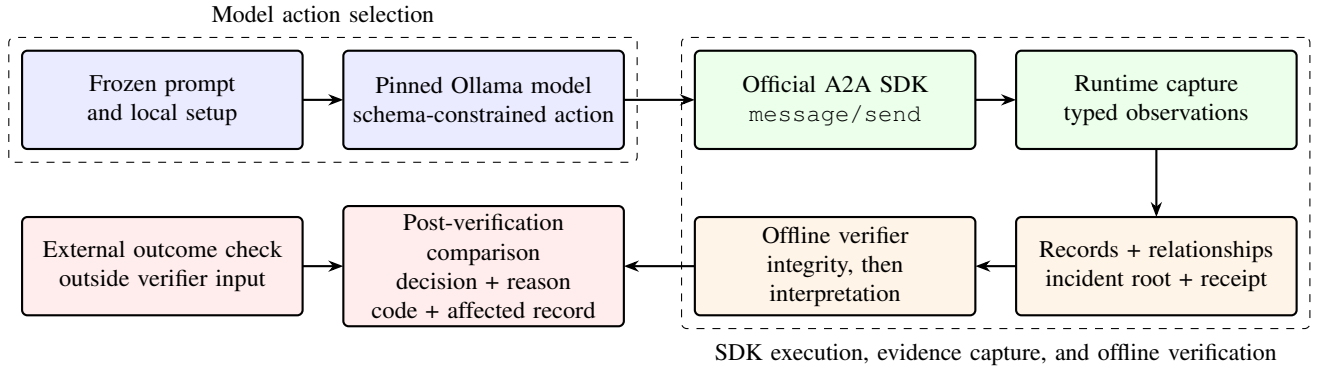

Fig.~\ref{fig:promptvalidation} shows the execution flow and its information boundary. The model receives the task input and permitted actions, but it does not receive the condition/control label, expected finding, or external outcome. The offline verifier receives the preserved bundle, receipt, profile, and trusted signer--key binding. After verification, the harness compares the decision with the external outcome. For condition runs, it also compares the affected-record identifier with the external target.

\subsection{Scenario Taxonomy and Evaluation Tasks}
The six attack-informed condition families were selected from the protocol-aware taxonomy in~\cite{li_a2asecbench_2026}. They evaluate forensic conditions in preserved evidence without reproducing live exploits or measuring attack success. Each condition is expected to return the listed finding, while its matched control is expected to return \code{FORENSICALLY_CONSISTENT}.

\begin{table*}[t]
\centering
\caption{Evidence and condition findings for six LLM-backed condition--control pairs.}
\label{tab:taxonomy}
\fontsize{9}{11}\selectfont
\setlength{\tabcolsep}{4pt}
\renewcommand{\arraystretch}{1.05}
\begin{tabularx}{\textwidth}{@{}L{0.405\textwidth} L{0.315\textwidth} Y@{}}
\toprule
Family: condition / matched control &
Preserved evidence and relations &
Condition finding \\
\midrule
\textbf{AgentCard spoofing:}
selected card: lookalike / trusted exact match &
Candidate, trusted, selected cards;
discovery and endpoint edges &
\decisioncode{LOOKALIKE_AGENTCARD}{SELECTED} \\
\addlinespace[2pt]
\textbf{Capability cloaking:}
invoked capability: undeclared / declared &
Selected card, invocation;
artifact-production edges &
\decisioncode{OBSERVED_CAPABILITY}{NOT_DECLARED} \\
\addlinespace[2pt]
\textbf{Cycle overflow:}
repeated route without progress / bounded route with progress &
Routing events, exchanges;
precedence edges &
\decisioncode{NON_TERMINATING}{ROUTING_CYCLE} \\
\addlinespace[2pt]
\textbf{Half-open task flooding:}
saturated input-required tasks / tasks completed after follow-up &
Task-load snapshots, transitions;
slot-occupancy edges &
\decisioncode{HALF_OPEN_TASK}{SATURATION} \\
\addlinespace[2pt]
\textbf{ASRF:}
dereferenced URI: disallowed / allowed &
URI reference, dereference, response;
causal edges &
\decisioncode{POLICY_DISALLOWED}{URI_DEREFERENCE} \\
\addlinespace[2pt]
\textbf{ATSI:}
artifact rendered without / after sanitization &
Artifact, render, sanitizer
records and relations &
\decisioncode{ACTIVE_CONTENT}{WITHOUT_SANITIZATION} \\
\bottomrule
\end{tabularx}
\end{table*}

\textbf{Matched-control construction and action selection.}
Within each family, the condition and control use the same task type, pinned model, permitted action set, and official-SDK route. The frozen manifest changes only the rule-triggering input or local setup and does not preselect the action. In both cases, the model returns one schema-constrained action, which the harness validates and applies before SDK execution. For ASRF, both cases request retrieval of the supplied URI. The condition uses a profile-disallowed URI and the control uses an allowed URI. Both requests are redirected to controlled test services running on the same computer. These services return controlled responses, and no external destination is contacted.

A marker-based propagation check inserts a unique inert marker into a controlled input and tests whether the same marker appears in the corresponding trial-local URI-service or renderer log. It confirms marker arrival at the observed surface, not script execution or exploit success.

\textbf{Outcome separation.}
The harness records an external reference outcome outside the preserved trace and excludes it from both model and verifier input. After verification, the harness compares the verifier decision with that external outcome. For each condition case, it also compares the reported affected-record identifier with the external target. Agreement indicates that the verifier produced the expected decision and localized the finding to the expected record. Matched controls have no expected affected record and are not scored for localization.

\textbf{Timing checks.} The separate verifier-scaling experiment uses valid bundles with exactly 10, 50, 100, 250, 500, 805, and 1,000 record-plus-relationship commitments. Larger fixtures add profile-valid artifact records and their exchange-to-artifact relationships, while one auxiliary AgentCard record is used to obtain odd leaf counts. Each size is measured in five randomized batches with ten untimed warm-ups and 100 timed verifications per batch. Trace construction, receipt signing, SDK execution, network activity, and file loading are excluded.

\textbf{Official-SDK execution path.} Each run discovers an AgentCard and submits a JSON-RPC \code{message/send} request through \code{a2a-sdk==1.1.0}. The request executes through the SDK server before the harness preserves SDK-facing objects and linked profile-specific observations. The bundle and receipt are then reloaded for offline verification against the external trusted binding. 

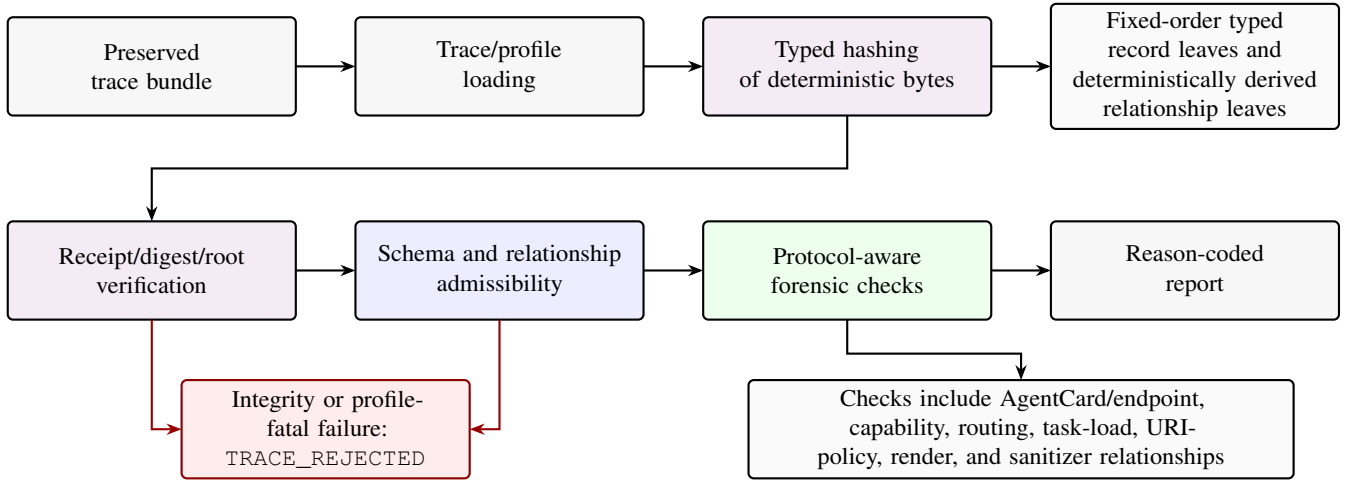
\begin{figure*}[!t]
\centering
\begin{tikzpicture}[
  main/.style={
    draw, rounded corners=2pt, thick, align=center,
    fill=gray!6, minimum height=13mm,
    text width=36mm, font=\fontsize{9}{11}\selectfont,
    inner sep=3pt, outer sep=0pt
  },
  commit/.style={main, fill=violet!8},
  gate/.style={main, fill=blue!7},
  check/.style={main, fill=green!7},
  reject/.style={
    main, draw=red!55!black, fill=red!7
  },
  detail/.style={
    main, fill=gray!4
  },
  arrow/.style={-{Stealth[length=2.0mm]}, thick},
  rejectarrow/.style={
    -{Stealth[length=2.0mm]}, thick, draw=red!60!black
  }
]

\node[main] (b1) at (0,0)
  {Preserved\\trace bundle};
\node[main] (b2) at (4.60,0)
  {Trace/profile\\loading};
\node[commit] (b3) at (9.20,0)
  {Typed hashing\\of deterministic bytes};

\node[commit] (b4) at (0,-2.70)
  {Receipt/digest/root\\verification};
\node[gate] (b5) at (4.60,-2.70)
  {Schema and relationship\\admissibility};
\node[check] (b6) at (9.20,-2.70)
  {Protocol-aware\\forensic checks};
\node[main] (b7) at (13.80,-2.70)
  {Reason-coded\\report};

\draw[arrow] (b1) -- (b2);
\draw[arrow] (b2) -- (b3);
\draw[arrow] (b3.south) -- (9.20,-1.35) -| (b4.north);
\draw[arrow] (b4) -- (b5);
\draw[arrow] (b5) -- (b6);
\draw[arrow] (b6) -- (b7);

\node[detail] (d1) at (13.80,0)
  {Fixed-order typed record leaves and\\
   deterministically derived relationship leaves};
\draw[arrow] (b3.east) -- (d1.west);

\node[reject] (d2) at (2.30,-4.80)
  {Integrity or profile-fatal failure:\\
   \texttt{TRACE\_REJECTED}};
\draw[rejectarrow] (b4.south) |- (d2.west);
\draw[rejectarrow] (b5.south) |- (d2.east);

\node[detail, text width=70mm] (d3) at (11.50,-4.80)
  {Checks include AgentCard/endpoint, capability,
   routing, task-load, URI-policy, render, and sanitizer relationships};
\draw[arrow] (b6.south) -- (9.20,-3.80) -| (d3.north);

\end{tikzpicture}

\caption{Verification pipeline. Receipt, capture-digest, root and profile-admissibility checks precede interpretation. Fatal failures return
\code{TRACE_REJECTED}.}
\label{fig:verifier}
\end{figure*}

Fig.~\ref{fig:verifier} summarizes the integrity and profile-admissibility checks that precede forensic interpretation.

\subsection{Results}
The primary LLM-backed matrix contained 240 official-SDK executions. All 120 condition runs produced the intended observation and returned the expected finding (descriptive 95\% Wilson interval: [0.969, 1.000]). None of the 120 matched controls produced an indication ([0, 0.031]), and all 240 incident roots and Ed25519 receipts verified. The affected-record identifier matched the external target in all 120 condition runs.  Across the 240 primary scenario measurements, median in-memory latency of the full offline verifier was 1.61~ms and the 95th percentile (p95) was 15.64~ms. For the scenario measurements, p95 uses linear interpolation at rank $1+0.95(n-1)$ in the sorted sample, where $n$ is the sample size.

Table~\ref{tab:familyresults} reports correctness, trace size, and verification latency for each family.

\begin{table}[t]
\centering
\caption{Per-family results from the 240 preserved scenario traces.}
\label{tab:familyresults}
\footnotesize
\setlength{\tabcolsep}{1.5pt}
\renewcommand{\arraystretch}{1.1}
\begin{tabularx}{\columnwidth}{@{}>{\raggedright\arraybackslash}Xrrrrrr@{}}
\toprule
Family & \shortstack{Condition\\findings} & \shortstack{Control\\indications} & \shortstack{Correct\\localizations} & \shortstack{Median\\size\\(kB)} & \shortstack{Median\\(ms)} & \shortstack{p95\\(ms)} \\
\midrule
AgentCard spoofing & 20/20 & 0/20 & 20/20 & 20.71 & 1.78 & 3.52 \\
Capability cloaking & 20/20 & 0/20 & 20/20 & 10.25 & 1.01 & 1.70 \\
Cycle overflow & 20/20 & 0/20 & 20/20 & 40.92 & 2.18 & 5.34 \\
Half-open task flooding & 20/20 & 0/20 & 20/20 & 195.69 & 12.33 & 26.39 \\
ASRF & 20/20 & 0/20 & 20/20 & 8.81 & 1.06 & 2.21 \\
ATSI & 20/20 & 0/20 & 20/20 & 9.42 & 1.08 & 2.06 \\
\bottomrule
\end{tabularx}
\par\smallskip
\begin{minipage}{\columnwidth}
\footnotesize
Size and latency summarize all 40 traces per family. Size is the stored
\code{forensic_capture.json} file size, including JSON formatting and
excluding the separate receipt. Here, 1~kB = 1,000 bytes. Latency is in-memory
full-verifier time. Localizations refer to the 20 condition runs.
\end{minipage}
\end{table}

All model decisions recorded in the 240 preserved final runs were schema-valid. Within each family and run type (condition or control), all 20 runs selected the same action. Condition and control actions differed in five families. For ASRF, all 40 runs selected URI retrieval. The supplied destination determined whether the policy condition was present.

The three configurations produced 720 results from the same 240 preserved LLM-backed traces. The root-only configuration recorded \code{INTEGRITY_}\allowbreak\code{VERIFIED} as an internal status for 240/240 executions, confirming trace integrity without a forensic decision. The records-only configuration, with relationship use disabled, returned \code{INCONCLUSIVE} for 240/240 executions. The full verifier agreed with the external outcome check in all 240 executions, classifying 120 as indicated and 120 as consistent.

\begin{figure}[t]
\centering
\resizebox{0.97\columnwidth}{!}{
\makebox[0.97\columnwidth][c]{%
\begin{tikzpicture}[
  font=\fontsize{9}{11}\selectfont,
  x=0.9cm, y=1cm,
  inner sep=1pt,
  line cap=round, line join=round
]
\def\plotw{7.4}
\def\ploth{4.0}

\foreach \y/\label in {0/0,0.8/10,1.6/20,2.4/30,3.2/40} {
  \draw[gray!25, line width=0.25pt] (0,\y) -- (\plotw,\y);
  \draw (-0.06,\y) -- (0.06,\y);
  \node[anchor=east] at (-0.12,\y) {\label};
}
\foreach \x/\label in {0/0,1.48/200,2.96/400,4.44/600,5.92/800,7.4/1000} {
  \draw[gray!20, line width=0.25pt] (\x,0) -- (\x,\ploth);
  \draw (\x,-0.06) -- (\x,0.06);
  \node[anchor=north] at (\x,-0.12) {\label};
}

\draw[black, line width=0.55pt] (0,0) -- (\plotw,0);
\draw[black, line width=0.55pt] (0,0) -- (0,\ploth);
\node[anchor=north] at (3.7,-0.72)
  {Committed leaves};
\node[rotate=90, anchor=south] at (-0.98,2.0)
  {Offline-verification latency (ms)};

\draw[blue!70!black, line width=1.05pt]
  (0.074,0.0364) --
  (0.370,0.1000) --
  (0.740,0.1860) --
  (1.850,0.4598) --
  (3.700,0.9692) --
  (5.957,1.9031) --
  (7.400,2.4680);
\foreach \x/\y in {
  0.074/0.0364,0.370/0.1000,0.740/0.1860,1.850/0.4598,
  3.700/0.9692,5.957/1.9031,7.400/2.4680
} {
  \fill[blue!70!black] (\x,\y) circle (1.65pt);
}

\draw[red!78!black, line width=1.0pt, dashed]
  (0.074,0.0723) --
  (0.370,0.1531) --
  (0.740,0.3106) --
  (1.850,0.7373) --
  (3.700,1.3853) --
  (5.957,3.0425) --
  (7.400,3.7149);
\foreach \x/\y in {
  0.074/0.0723,0.370/0.1531,0.740/0.3106,1.850/0.7373,
  3.700/1.3853,5.957/3.0425,7.400/3.7149
} {
  \fill[red!78!black] (\x,\y) rectangle ++(0.10,0.10);
}

\draw[blue!70!black, line width=1.05pt] (0.35,3.72) -- (0.82,3.72);
\fill[blue!70!black] (0.585,3.72) circle (1.65pt);
\node[anchor=west] at (0.95,3.72) {Median};
\draw[red!78!black, line width=1.0pt, dashed] (0.35,3.15) -- (0.82,3.15);
\fill[red!78!black] (0.535,3.10) rectangle ++(0.10,0.10);
\node[anchor=west] at (0.95,3.15) {p95};
\end{tikzpicture}%
}%
}
\caption{In-memory latency of the full offline verifier by bundle size (500 runs per size).}
\label{fig:scaling}
\end{figure}
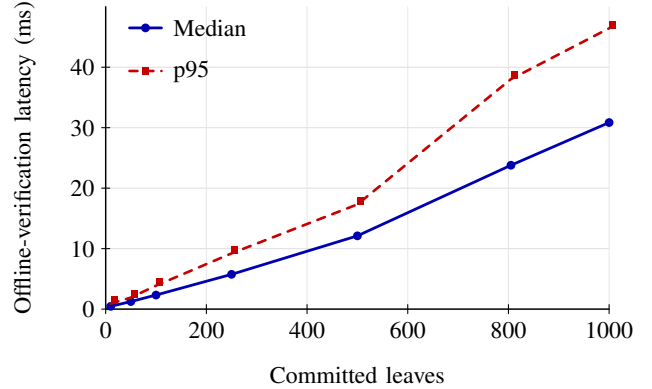

Fig.~\ref{fig:scaling} reports the separate verifier-scaling experiment. All 3,500 timed verifications returned the expected consistent decision, verified the incident root, and verified the Ed25519 receipt. From 10 to 1,000 committed leaves, the median increased from 0.46 to 30.85~ms and p95 from 0.90 to 46.44~ms. The monotonic increase characterizes the implemented verifier under the evaluated record--relationship composition.

\section{Discussion and Limitations}
\label{sec:discussion}
Within the evaluated profile, the results distinguish integrity verification from forensic interpretation. Root-only verification confirmed integrity for all 240 traces but produced no forensic conclusion, while records-only verification remained \code{INCONCLUSIVE}. The full verifier matched the external reference outcome for all 240 traces: all 120 condition runs produced the expected findings, and all 120 matched controls remained \code{FORENSICALLY_CONSISTENT}. Our design stores the derived relationships explicitly and protects them with cryptographic commitments. Because these relationships can be derived from the same records, this comparison does not show that separate relationship commitments are necessary for offline interpretation. Verification cost increased monotonically with commitment count, with median latency rising from 0.46~ms at 10 leaves to 30.85~ms at 1,000 leaves.

A2A-ForensicTrace supports post-incident analysis under an explicit verifier profile. The integrity claim applies after signing. It does not cover omitted events, compromised capture, internally consistent false records, or compromise of the signer, verifier, or trust store. A valid receipt binds a trusted key to a committed trace. It does not establish why an action occurred (intent) or identify the responsible actor or system (attribution). It does not prove that all relevant events were captured (completeness). It also does not prove that the captured observations reflect the original execution (capture truth). To check for substitution by an older valid trace, investigators can compare the receipt with an independently trusted record of the expected incident and root. The signed issue time alone does not show that the trace is the latest expected version. Operational deployments should preserve historical signer--key bindings across key rotations and retain the dated revocation information and trust policy used at verification time. Without updated revocation information, offline verification cannot determine whether a key was revoked after the last available status update. These freshness and key-lifecycle procedures, legal chain of custody, trusted timestamps, cross-agent clock synchronization, and evidentiary admissibility are outside the evaluated scope.

The evaluation used \code{a2a-sdk==1.1.0} and covered only its non-streaming JSON-RPC \code{message/send} compatibility path. It used one LLM-backed execution matrix, one profile version with six family-specific manifests, one pinned local model, frozen inputs, and a trial-local loopback testbed. The evaluation does not cover streaming, push notifications, retries, idempotency, alternate bindings, or the full A2A lifecycle. ASRF uses trial-local loopback mapping, and ATSI uses an inert marker with a non-browser renderer. The study tests end-to-end evidence handling when model action precedes SDK execution. It is not an LLM-safety or attack-success benchmark. The six A2ASecBench categories do not cover all A2A forensic conditions~\cite{li_a2asecbench_2026}.

The Wilson intervals summarize the fixed run schedule and do not support population-level generalization. Timing and scaling results are specific to the evaluated environment, procedures, and valid-bundle composition. The reported evaluation covers predefined scenarios and valid bundles. Systematic post-capture mutation tests and independently authored scenarios remain outside its scope.

\subsection{Practical Implications}
A2A-ForensicTrace helps investigators review incidents using preserved A2A runtime evidence. With the trace bundle, signed receipt, verification profile, and trusted signer-key binding, investigators can verify integrity and apply forensic checks offline. Reason codes and record identifiers point to the evidence supporting each finding. Organizations should retain these verification inputs and control access to the preserved evidence.

\section{Conclusion}
\label{sec:conclusion}
This paper presented A2A-ForensicTrace, which commits selected A2A runtime records and derived relationships, then applies profile-bound checks only after integrity and admissibility succeed. Across 240 LLM-backed official-SDK executions, all 120 condition runs produced the expected bounded finding. None of the 120 matched controls produced an indication, and every incident root and Ed25519 receipt verified. Within the implemented profile, the results support offline integrity verification of selected A2A runtime observations preserved in a tamper-evident incident trace. The verifier uses committed records and relationships to produce bounded protocol-level findings. Future work will broaden A2A lifecycle coverage and study reduced-disclosure evidence bundles that preserve the required offline verification decision while revealing fewer records and relationships. 

\bibliographystyle{IEEEtran}
\bibliography{Paper3B}
\end{document}